\documentclass[12pt,preprint]{aastex}

\shorttitle{Mixing in Stars}
\shortauthors{Young et al.}

\newcommand{\sol}{$M_\odot$}
\def \nuc#1#2{\relax\ifmmode{}^{#1}{\protect\text{#2}}\else${}^{#1}$#2\fi}

\received{2002 August 28}
\begin{document}

\title{Stellar Hydrodynamics in Radiative Regions}

\author{Patrick A. Young, Karen A. Knierman, Jane R. Rigby, and David Arnett}
\affil{Steward Observatory, University of Arizona, 
933 N. Cherry Avenue, Tucson AZ 85721}
\email{payoung@as.arizona.edu, kknierman@as.arizona.edu, 
jrigby@as.arizona.edu, darnett@as.arizona.edu}

\begin{abstract}

We present an analysis of the response of a radiative region to
waves generated by a convective region of the star; this wave treatment
of the classical problem of ``overshooting'' gives
extra mixing relative to the treatment traditionally used in stellar
evolutionary codes. The interface between convectively
stable and unstable regions is dynamic and nonspherical, so that the
nonturbulent material is driven into motion, even in the absence of
``penetrative overshoot.'' These motions may be
described by the theory of nonspherical stellar pulsations, and are
related to motion measured by helioseismology. Multi-dimensional numerical
simulations of convective flow show puzzling features which we explain
by this simplified physical model. Gravity waves generated at the
interface are dissipated, resulting in slow circulation and
mixing seen outside the formal convection zone. The approach may
be extended to deal with rotation and composition gradients
(``semiconvection'').

Tests of this description in the stellar evolution code TYCHO produce carbon
stars on the asymptotic giant branch (AGB), an isochrone age for the
Hyades and three young clusters with lithium depletion ages from brown
dwarfs, and lithium and beryllium depletion consistent with
observations of the Hyades and Pleiades, all without tuning
parameters. The potential insight into the different contributions of
rotational and hydrodynamic mixing processes could have important
implications for realistic simulation of supernovae and other
questions in stellar evolution.

\end{abstract}

\keywords{stars: evolution 
- stars: fundamental parameters - hydrodynamics 
- convection - lithium}

\section{INTRODUCTION}

The nature of mixing in stars is a perpetual problem in stellar
evolution. The standard mixing length theory of convection 
\citep{kippen} is remarkably effective for a
one-dimensional, ensemble average of convective energy
transport. However, observations seem to indicate that more mixing
occurs in stars than is expected. For example, measurements of the 
apsidal motion
of binary star orbits give a measure of the density structure of the
components. Comparisons with mixing-length models indicate that real
stars have larger convective cores than predicted by theory
\citep{ymal01}. Models of double-lined, eclipsing binaries with well
determined masses and radii also require additional mixing to match
observations \citep{ymal01, pol97, rib00}. Determinations of young
cluster ages independent of isochrone fits to the main sequence using
the lithium depletion edge in brown dwarfs give substantially older
ages which can also be reconciled by increased mixing \citep{ssk98}.

Parameterized descriptions of
mixing can tell us a great deal, but only in well populated regions of
the H-R diagram where high-quality observational constraints are
numerous. Light element depletion on the pre-main sequence (pre-MS)
and convective core sizes, and thus lifetimes and luminosities, on the
main sequence are affected \citep{pt02, ymal01}. For low and
intermediate mass stars s-process nucleosynthesis on the AGB, ISM
enrichment, and white dwarf sizes and compositions are strongly
influenced \citep{wk98}. In massive stars the size of the heavy
element core and mixing in the high-temperature burning shells may
substantially impact supernova nucleosynthesis and explosion
mechanisms (BA98,AA00).

It has long been known \citep{spi67} that mixing-length theory, by
approximating derivatives poorly, must have problems at the interface
between convective and nonconvective regions, posing an embarrassment
for stellar evolution. \citet{ss65} discussed the problem using laminar
hydrodynamic theory in the convection zone, which ignores the strongly
turbulent nature of stellar convection. \citet{ss73} examined the
ballistics of a convective blob; this particle approach does not
impose continuity (mass conservation) on the dynamics. 
These two approaches are well represented in the extensive literature
on the subject. Most modern stellar evolution codes seem to use
either mixing length theory \citep{kippen,clay83} 
or the full spectrum theory \citep{cm91} in the turbulent regions
and assume other regions are static.

We note that the correct equations
for describing stellar {\em nonconvective} regions are hydrodynamic,
not static \citep{cox80}. 
If symmetry is broken, as by perturbations from a convective region,
these motions are also three-dimensional.
We examine the hydrodynamics
induced in radiative regions, due to the fact that the 
convective/nonconvective interface is neither static nor spherical.
The problem becomes one of driven, non-radial, non-adiabatic
pulsations \citep{hansen}.
Numerical simulations (especially \citet{ba98} BA98, and \citet{sa00}
AA00)
lead us to the possibility that large wavelength modes, specifically
plumes, are dominant in the coupling at this interface. We
suggest how this coupling works, and how it necessarily implies
a slow mixing into the radiative region.
We show how a simple version of this picture may be implemented
in a stellar evolutionary code. Our theory is complementary to 
theories of the turbulent convective region, such as the standard
mixing length theory \citep{kippen} or the full spectrum theory
\citep{cm91}. 

In this paper we focus on the simplest case, and provide a
lower limit on ``extra'' mixing --- which is above that obtained with
conventional stellar evolution theory.
We argue that our theory is a necessary part of a
complete solution, but believe that other aspects of hydrodynamics (such as
rotation \citep{mm89, kq97, every, ct99, pin02, tkz02}) 
are also important, and may be synthesized
into a more complete theory.
We present several tests of our approach by comparison with observations.

\section{Implications of Numerical Simulations}

We begin by examining multidimensional numerical simulations, which
are nonlocal and fully nonlinear.
We have been most influenced by BA98 and AA00, but
have also studied \citet{pw01} and \citet{bt02} in some detail.
These simulations give us a glimpse
of the hydrodynamic behavior of the interface of the convection
zone, from which we can begin to construct a theoretical picture. 
These simulations are not entirely accurate
descriptions of the star due to the limited range of resolution.  
Turbulent structure is expected to span all size scales
down to the local diffusion scale, which is much smaller than the
resolution element of any simulation which captures the large 
scale structure. Further impact of sub-resolution
scale physics is discussed in \citet{can00}. 
The maximum Reynolds number of the simulations (in 3D)
to $R \sim 10^8$, whereas in stars $R$ may be as high as
$10^{14}$. Turbulence may become
completely chaotic at $R >> R_0$, a regime which is not amenable to
exploration numerically or experimentally \citep{dim01}. While
microscopic mixing is not well treated because it is dominated by
processes with length scales smaller than the resolution of the
simulations, energy and bulk transport are dominated by
processes with large length scales, and may be modeled better. It is
necessary to develop a theoretical understanding of the
processes involved, rather than relying entirely upon numerical results.

These simulations presented us with two puzzles:
\begin{itemize}
\item Large density perturbations appear at the interface between
convective and nonconvective regions (BA98).
\item Slow vortex motion appears outside the formally convective
region, giving a slow mixing (AA00).
\end{itemize}
How can this be understood?

\subsection{Convective forcing}

Inside a stellar convection zone, the velocity field has significant
vorticity; outside the velocity is assumed negligible. 
Hydrodynamically, this interface corresponds to a 
{\em surface of separation} between
rotational (${\bf \nabla \times v \neq 0}$) 
and irrotational flow (${\bf \nabla \times v = 0}$)  
\citep{ll59}, see \S 34.

To be specific we consider the outer edge of a convective oxygen burning shell
(BA98,AA00). This is a simple case in that it avoids the added
complexity of a photospheric boundary condition (as opposed to
simulations of the solar convection zone, for example \citet{cd02,fls96})
and can be evolved numerically on the evolutionary timescale, since
the nuclear and sound-travel timescales are commensurate. The convection
does work on the interface between laminar and turbulent regions, with
a total power (luminosity)
\begin{equation}
L_{conv} = A \delta P v = A P v_s (\delta P/P)(v/v_s),
\end{equation}
where $A=4 \pi r^2$ is the spherical area, $v_s$ is the sound speed,
$v$ the transport velocity by convection, and $\delta P$ the
pressure fluctuation. Inserting numerical values from the simulations
we find
\begin{equation}
(\delta P/P)(v/v_s) = 10^{-4}
\end{equation}
and since $(\delta P/P) \simeq (v/v_s)$, we have a Mach number of
\begin{equation}
(v/v_s) = 10^{-2}.
\end{equation}
This estimate uses mathematical relations in the spirit of mixing
length theory, and gives an average velocity. Examination of the
numerical results shows that the actual velocity is concentrated in
plumes which occupy a smaller cross-sectional area, but have higher
speeds \citep{hurl96}. 
There are significant density perturbations at the boundary
between laminar and turbulent flow. This is sufficient to drive a
nontrivial acoustic flux and cause significant non-radial density
perturbations $\delta \rho / \rho$ of a few percent 
(see BA98, figure~3 and figure~7 and AA00, figure~8).

For earlier and less vigorous burning stages, the Mach number is
smaller, so that neglect of acoustic flux may not be so atrocious.
However, these stages are also longer, so that the accumulated effect
of the waves may still be significant.
These enhancemented density variations at the interface are a robust 
feature in simulations; three-dimensional calculations of
the solar convective zone and red giant stars have displayed similar
pumping of gravity waves \citep{bt02,pw01}.

In the stellar interior, convective luminosity may be estimated 
without any detailed theory of convection. The hydrodynamic motion is
nearly adiabatic, so the radiative flux is close to that for 
radiative diffusion for an adiabatic temperature gradient. 
The total luminosity is determined
from the conservation of energy, so that the convective luminosity is
the difference $ L_{conv} = L_{total} - L_{rad}$ \citep{kippen}.
At the edge of the convective region, we identify this with the
energy flux available to drive waves by the decceleration of plumes.
The precise fraction of the luminosity that goes into driving
depends upon the detailed physics of the convective interface
\citep{gmk94}; our simulations suggest the kinetic part is comparable
to the thermal part of the convective flux (BA98, figure~3).

\subsection{Hydrodynamic response}
What does this do to the radiative region? The natural modes for 
nearly laminar
flow are irrotational, and in general will be incommensurate with the
rotational flows of the convective zone.  There will be a mismatch at
the boundary, so that the boundary matter will be driven, exciting
waves. Because the motion of the plumes is generally subsonic, the
coupling will be biased toward g-modes, which have longer
periods. \citet{sa00} find a combination of waves, having both g-mode
and p-mode character \citep{cowl41}. The waves exhibit an exponential
fall-off moving away from the interface into the radiative region,
but also significant compressible effects (density fluctuations).
See AA00, figures~8, 11-14 for detail.

\begin{figure}
\figurenum{1}
\plotone{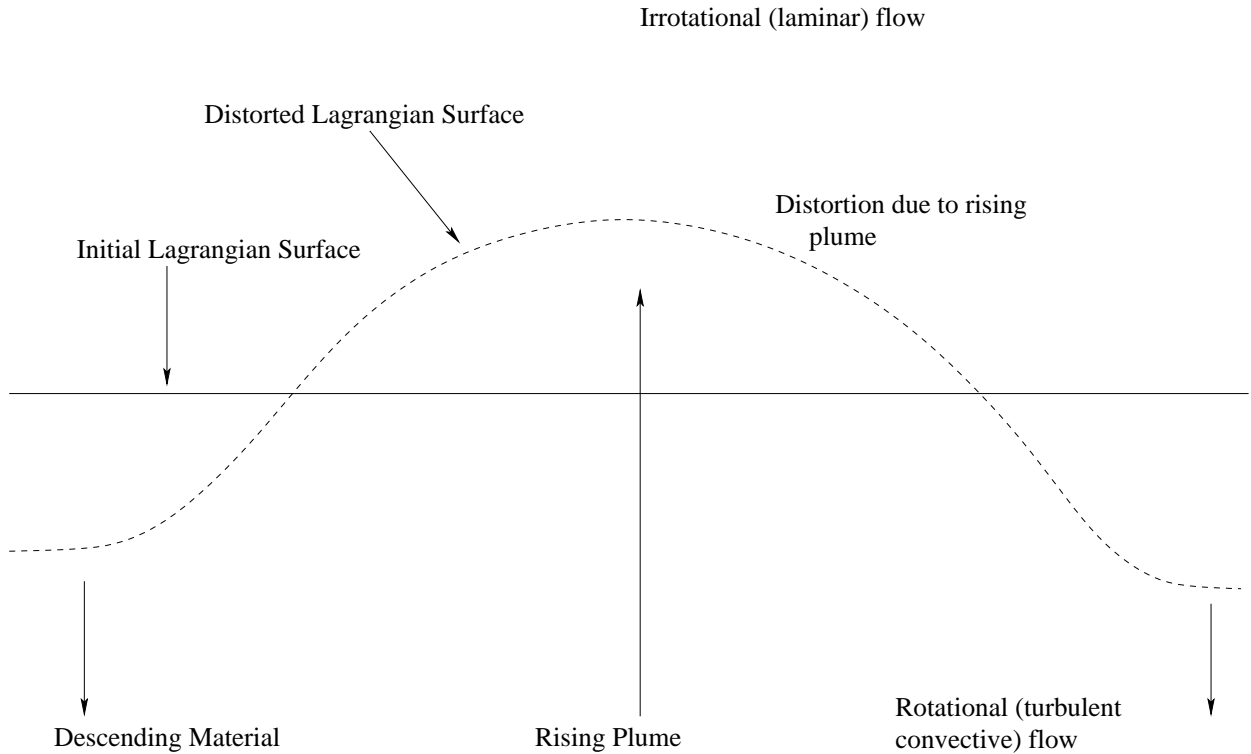}
\caption{Lagrangian (co-moving) fluid surfaces at boundary between
convectively stable and unstable regions. The distortion due to a
rising plume and a downdraft are shown. The original spherical
boundary is distorted as interface material bobs up and down,
generating gravity waves. Contrary to the usual assumption in stellar
evolution simulations, the convective boundary is neither spherical
nor static. Dissipation of the gravity waves actually causes slow
circulation in the nominally laminar region outside the convective
zone.}
\end{figure}
\placefigure{fig1}

Figure 1 shows a schematic of the behavior of the interface of the
convective region. Three dimensional simulations of convection also
show up-down asymmetry \citep{pw01, bt02}. For the solar convective
envelope, with driving caused by the entropy decrease from radiative loss
near the photosphere, plumes tend to move downward from the
photospheric surface. For oxygen shell burning, plumes tend to move
upward from the burning shell, in which nuclear energy release causes
an entropy increase. Neutrino cooling tends to cause plumes that are
directed downward (BA98).

As a plume encounters a boundary, it pushes the over(under)lying
material, distorting the boundary. Part of the plume's kinetic energy
goes into raising the potential energy of the displaced region. When
the plume stalls, this potential energy is converted into motion in
the opposite direction of the plume's velocity. Gravity waves are
generated. Contrary to the usual assumption in stellar simulations,
the convective boundary is neither spherical nor static. This resolves
a paradox of mixing length theory, in which the convective velocity
has a discontinuity at the convective boundary, going from a finite
value to zero. 

The surface of separation is a nonspherical comoving (Lagrangian) boundary,
which moves relative to the spherical (Eulerian) boundaries of a
stellar evolution code. The spherical shells do move on average with
the matter, in that they may contain a fixed amount of mass, but it
need not be the same matter. While the interface moves across
a spherical shell, it may later move back. 
Motion does not necessarily give mixing. 
Note that this goes beyond the usual notion of spherical Lagrangian
shells in a stellar evolutionary code; hydrodynamic motion is faster
than slow secular evolution, so that the spherical shells seem relatively
fixed in space (that is, Eulerian); see \citet{cox80}.

\subsection{g-modes}

To the extent that the time scale for heating and cooling the radiative region 
is longer than that for hydrodynamic motion, Kelvin's circulation theorem
holds \citep{ll59}. 
Further, if the hydrodynamic motion is slow (strongly subsonic), it
is described by a velocity potential, $ {\bf v} = \nabla \phi$, where
$\phi$ satisfies Laplace's equation $\nabla^2 \phi = 0$.
If we take a Cartesian coordinate system (x,y,z) with z positive along
the radial direction ${\bf r}$,  $z=0$ at the interface between
convection and nonconvection, and assume the waveform transerse to
z is periodic, then
\begin{equation}
\phi = f(z) e^{i(k_x x + k_y y - \omega t)} ,
\end{equation}
so
\begin{equation}
\nabla ^2 \phi = \phi ( -(k_x^2 + k_y^2) + {1 \over f} 
{\partial ^2 f \over \partial z^2 } )= 0 ,
\end{equation}
giving $ f = e^{\pm k z}$, where $k$ is the transverse wave number,
defined by $k^2 = k_x^2 + k_y^2$. The sign choice comes from the boundary
condition, so that the function decreases exponentially away from the
boundary  $z=0$. Waves of longer wavelength (small $k$) extend farther
from the boundary.
While this is a useful guide, the actual waves (AA00) are not strictly
incompressible (${\bf \nabla \cdot v} =0$; density variations occur, and
are important for damping the waves. 

We relate the wave number to the frequency by equating the acceleration
in the $z$ direction to the corresponding force per unit mass. For
an incompressible liquid it gives $\omega ^2 = k g$, while the compressible
case results in the Brunt-V\"ais\"al\"a frequency
\begin{equation}
N ^2 = {g \delta \over H_P}(\nabla_{ad} - \nabla + 
{ \varphi \over \delta}  \nabla_\mu),
\end{equation}
from \citet{kippen}, eq. 6.18, where the symbols have their usual meaning,
or \citet{hansen}, eq. 5.35 and 10.92.
Our system is finite, so only a discrete spectrum of waves is possible.
Notice than the quantity in parenthesis is the Ledoux condition for
convective instability, and has implications for regions with compositional
gradients, which we do not pursue here.

The longest wavelengths penetrate further, and will be most effective for
mixing. The maximum wavelength generated will
depend upon the details of the convective driving. 

With a complete theory of turbulent
convection we could simply determine a transfer function for the
excitation of waves in the radiative region \citep{gmk94}. 
Mixing length theory is the
simplest; it maintains that only one dominant wavelength need be
considered --- the mixing length. The Canuto-Mazzitelli theory 
gives a broader spectrum of modes but they peak in the same place
(see \citep{cm91}, Figure~1). For simplicity we take the appropriate
wavelength to be  equal to
the length scale we would derive from the size of the plumes seen
in simulations. 

\subsection{Dissipation of waves}

The driving of the waves must be balanced by their dissipation for
a steady state to result. In the stellar plasma this will usually be
due to thermal diffusion of radiation. Such dissipation will be 
faster at the shorter wavelengths; for a given amplitude they have
the largest gradients.
For a given wave, we could integrate the wave equation \citep{cox80}
for a precise result. The precision would be illusory in that the
range of relevant wavelengths would depend upon our ignorance of
the properties of the convective driving. Instead we give a simplier
example to illustrate the physics and make a preliminary estimate
of the importance of the process.

The canonical picture of damping of gravity waves is by viscosity
( \citet{ll59}, \S 25). Using the viscosity of a plasma in the absence
of magnetic fields (\citet{spz62}, \S 5.5), we find a damping time
of many gigayears for stellar conditions, so this is not the relavant
damping. The compressible effects give rise to temperature fluctuations;
this gives a pressure perturbation which resists the wave motion,
analogous to damping of stellar pulsations \citep{cox80}.
Following \citet{kq97}, the local radiative dissipation of gravity
waves is
\begin{equation}
\gamma \approx { 2 F_r k_r^2 H_T \over 5 P },
\end{equation}
where $F_r$ is the radiative flux at radius $r$ from the center of the
star, $k_r \approx N [l(l+1)]^{1 \over 2} / [ r \omega ]$ is the wave's
radial wave number for frequency $\omega$, $P$ is the pressure, and 
$H_T$ is the temperature scale height.

\subsection{Circulation and mixing}

A difficult step is the connection between the multidimensional flow
and the microscopic mixing. We argue that dissipation drives
circulation, which is likely to be turbulent. For the purposes of a
stellar evolution code we identify this with a diffusive velocity
$u_{k}(\Delta r)$, even though the physical identity is not exact. The
characteristic scale is $ l_{turb}$, and is determined from
simulations.

As we saw above, the coupling of convective plumes with the region of
laminar flow outside the convective region generates significant
density anisotropies and waves at the boundary. These low Mach
number waves can be described approximately
as  potential flow which we assume to
be dissipated over a distance determined by the hydrodynamics. This
damping is an entropy-generating process, causing vorticity which
allows for microscopic mixing of the material and slow circulation of
the mixed material well beyond the convectively neutral boundary.
Qualitatively, this is like breaking of wave crests on a sea.

For didactic purposes we will derive the generation of vorticity by
damping of the potential flow in a simple plane parallel
case. Following \citet{ll59}, \S 9,
\begin{equation}
\case{d{\bf v}}{dt} + {\bf v \cdot} \nabla {\bf v} = \nabla{\bf w} - T \nabla S
\end{equation}
 Discarding ${\bf v \cdot \nabla v}$ as small,
\begin{equation}
\nabla \times \case{d{\bf v}}{dt} = \nabla \times \nabla{\bf w} - \nabla \times (T \nabla S) 
\end{equation}
The term $\nabla \times \nabla{\bf w} \rightarrow 0$, giving
\begin{equation}
\case{d \nabla \times {\bf v}}{dt} = -T(\nabla \times \nabla S) + \nabla S \times \nabla T
\end{equation}
Discarding $-T(\nabla \times \nabla S) \rightarrow 0$ gives the final
form for the generation of vorticity,
\begin{equation}
\case{d \nabla \times {\bf v}}{dt} = \nabla S \times \nabla T
\end{equation}
In a perfectly spherically symmetric star $\nabla S \times \nabla T$\
would go to zero in the laminar regions. When we introduce
perturbations from the damping of the waves, however, we gain a cross
term which makes the time derivative of the vorticity non-zero. We
employ a standard style of perturbation analysis {\it a la}
\citet{ll59} or \citet{hansen}, discarding terms of higher than first order, and
examine the contribution from
\begin{equation}
\case{d \nabla \times {\bf v}}{dt} = \nabla S' \times \nabla T_0
\end{equation}
where $X'$ denotes a perturbation and $X_0$ denotes the unperturbed
value of a variable. Henceforth we will change notation to $X = X_0$
for simplicity. In the simplified plane parallel case and ignoring
unecessary constants,
\begin{equation}
\case{d \nabla \times {\bf v}}{dt} = \case{\partial S'}{\partial x}\times \case{\partial T}{\partial z}
\end{equation}
From the standard equations of stellar structure \citep{kippen} we take
\begin{equation}
\case{\partial T_0}{\partial z} = -\case{3\kappa \rho L}{16\pi ac r^2 T^3} \hat{z}
\end{equation}
and from thermodynamics \citep{r65}
\begin{equation}
\case{\partial S'}{\partial x} = \case{1}{T + T'}(4aT^3\case{\partial T'}{\partial x} + \case{P}{\rho^2}\case{\partial \rho'}{\partial x})\hat{x}
\end{equation}
We will assume an adiabatic case, such that $\rho^{\gamma - 1}T = const$ and 
\begin{equation}
\case{\partial \rho'}{\partial x} = \case{1}{\gamma - 1}T'^{(\case{2 - \gamma}{\gamma - 1})}\case{\partial T'}{\partial x}
\end{equation} 
After some algebraic manipulation,
\begin{eqnarray}
\case{d \nabla \times {\bf v}}{dt} = \case{\partial S'}{\partial x}\times \case{\partial T}{\partial z} = \ \ \ \ \ \ \ \ \ \ \ \ \     \\
\nonumber -\case{3\kappa \rho L}{16\pi ac r^2 T^3}\case{1}{T+T'}(4acT^3 + \case{1}{\gamma - 1}T'^{(\case{2 - \gamma}{\gamma - 1})}\case{P}{\rho^2})\case{\partial T'}{\partial x} 
\end{eqnarray}
Integrating over $dt$\ with the damping described in \S 2.4 and a
reasonable approximation to the wavefunction gives an estimate of the
vorticity. Further using the curl theorem and integrating the
vorticity over the path of a fluid element gives an estimate of the
diffusion velocity at a given radius.

\section{Implementation in stellar evolution}

To implement this mixing in TYCHO, we treat the
mixing as a diffusion process with a diffusion coefficient
\begin{equation}
D = \case{u_{k}(\Delta r) l_{turb}}{3}
\end{equation} 
constructed from the terms discussed in the previous section.  This
treatment leaves one free parameter, $\case{l_{turb}}{H_p}$, the
dominant scale length of the turbulence near impact of the plumes with
the boundary. This quantity is directly related to the dominant wavelength of
the gravity waves driving the mixing. There is power at all scales in
the convective region. The power is flat or slightly rising from the
largest scales to the value we choose for our treatment and then
follows a power law consistent with Kolmogorov turbulence down to the
smallest resolved scales \citep{pw01}. We estimate
$\case{l_{turb}}{H_p}\sim 0.1 - 0.15$ using three-dimensional
numerical simulations \citep{pw01}. Traditionally, model fits have
been improved by introducing free parameters such as the alpha
prescription for overshooting \citep{mm89}. Clearly, we should not
have infinite freedom to introduce parameters. While the parameterized
approach has yielded extremely important results in terms of
understanding the extent of the extra mixing observed in stars, it
gives us little insight into the underlying physics and has limited
predictive power. By fixing this quantity using results from
multi-dimensional hydro calculations, we are attempting to construct a
physical picture of the mixing in the radiative region with minimal
variability in parameters. We prefer to constrain our theory by
terrestrial simulations and experiment rather than astronomical
observation. This should increase the predictive power of the
theory. Additional simulations are needed to explore the behavior of
this scaling in a wider variety of conditions appropriate to stellar
astrophysics. Cases where the pressure scale height is divergent or
much larger than the convective scale, for example in the small
convective core of the ZAMS sun are of particular interest.

A desirable property falls naturally out of this
treatment. Three-dimensional hydro simulations indicate that
boundaries with shallow changes in the adiabatic gradient should be
able to mix over wider ranges in radii \citep{bt02}. This should
result in more mixing for higher mass stars, and more mixing in
convective cores than in envelopes, which seems to be supported by
parametrized overshooting in previous work \citep{pol98, mm89}. This
treatment preserves this behavior, since the region over which the
gravity waves are dissipated is larger in the more isentropic
environment of core convection. Also, the higher convective velocities
in H burning cores than envelopes result in a higher gravity wave flux
and larger mixing region, and similarly more mixing in He cores than
H. The extra mixing occurs over a significant fraction of a pressure
scale height in core convection (compare with values of 0.4--0.6$H_p$\
in parametrized overshooting) and $\la$ 0.05--0.1$H_p$\ for envelope
convection.

\section{THE STELLAR EVOLUTION CODE TYCHO}

All stellar evolution calculations presented below were performed
using the TYCHO 1-D stellar evolution code discussed by \citet{ymal01}
but with substantial improvements in several areas. The equation of
state (EOS) has been updated to use a modified version of the
\citet{ts00} tabular electron-positron EOS. It has been further
modified to have appropriate coulomb corrections for the weak
screening case and a Debye interpolation for strongly coupled
plasmas. This agrees to within 2\% (and usually to less than 0.1\%)
with the EOS tested empirically by the OPAL project's high energy
density laser experiments \citep{ir96}. There are significant
deviations from our EOS only where the OPAL models do not account for
contributions from electron degeneracy pressure. The size of the
reaction network was increased to 175 nuclei, and is well populated
all the way up to the iron peak. The low temperature opacities have
been completely revised to use tables from \citet{af94}, and are
interpolated to serve for any metallicity between zero and five times
solar. The mass loss at low $T_{eff}$ has been updated to use the
modified Reimers formulation presented in \citet{bl95}, which results
in much higher mass loss rates on the AGB. Alternatively, low
temperature mass loss may be switched off entirely to examine purely
episodic mass loss on the AGB. An ADI operator split has been
implemented in the mixing algorithm so that nuclear reaction
calculations will be informed about the change in composition, and the
thermodynamic variables used in the EOS and mixing routines will
properly take into account energy input by burning and neutrino
cooling. The mixing is also now time-limited rather than
instantaneous. Additional refinements improving the numerical
convergence of the code and its convergence at small timesteps have
also been incorporated. Experiments have been performed which include
heavy element diffusion, and give unsurprising results, consistent with
solar models from \citet{bpb01}. The version of the code used in this
study (TYCHO 6.11) does not incorporate heavy element diffusion, as
such an examination is beyond the scope of the current discussion. It
is also useful to separate the effect of settling out from the
phenomenon being examined. The timescale for settling is sufficiently
long that for ages much less than than of the Sun, the effect should
be negligible. The two quantitative cases presented herein both have
ages less than $10^9$\ yr, and should not be affected.

\section{COMPARISONS WITH PREVIOUS WORK}

Remedies to the problem of mixing have until recently largely been
phenomenological. The mixing beyond the standard model is parametrized
and labeled as ``overshooting'' in convective cores and
``undershooting'' in convective envelopes, or more generically as
overshooting in both cases. The term has been taken by various groups
to encompass both penetrative convection beyond the formal boundary of
convective stability and slow compositional mixing. The most common
overshooting prescription is ``alpha-overshoot'', where compositional
mixing is arbitrarily extended some fraction of a pressure scale
height beyond the boundary of the formally convective region
\citep{mm89}. More recently, \citet{pol98} have devised a
parameterization based upon the superadiabatic excess of the boundary,
which has the advantage of being tied to the structure of the
star. Parameter fitting of this sort is valuable in constraining the
extent of the extra mixing by astronomical observation, but gives us
little insight into the physical nature of the process. Overshooting
based on rotational mixing has also been proposed. It has been
particularly useful in solving the problem of the lithium gap in F
stars. The blue side of the dip is reasonably well modelled by
rotation-driven meridional circulation \citep{del98, bo02, pt02,
pin02}. Recent work describing angular momentum transport by gravity
waves has shown considerable success in matching the red side of the
dip \citep{ct99, tkz02}.  Rotation looks likely to be an important
contributor to the solution of the mixing problem, but is probably not
the whole of the story \citep{mm89, pin02}. It is our intention to
avoid the use of the term ``overshooting'' entirely so as to be free
of its associated connotations.

In this paper we discuss non-rotational hydrodynamic contributions to
the mixing from gravity waves generated at the surface of separation
between the convective and laminar regions of a star. \citet{gls91}
attempt to assess the contribution to mixing of gravity waves at the
convective boundary in the particular context of lithium depletion in
F type stars in young clusters. They conclude that the mechanism
produces the proper mixing behavior, but requires a gravity wave flux
a factor of fifteen larger than given by simple estimates. This
problem is not insurmountable. They themselves point out that the
efficiency of converting kinetic energy of convective fluid elements
increases significantly if the downflows driving the waves are
concentrated into narrow plumes. Simulations show the filling factors
of these plumes are a few percent \citep{pw01, bt02}. In addition,
\citet{can02} argues that turbulent mixing in a stellar context is
likely to persist for a larger range of conditions. In \citet{gls91},
the extent of mixing was limited by comparing an unperturbed stellar
model to a laminar stability model. The critical Richardson number
$Ri(cr)$ for which turbulence may persist once established is a factor
of four larger than $Ri^{l}(cr)$ for the breakdown of an established
laminar flow. In addition, radiative losses weaken stable
stratification and the gravity waves themselves act as an energy
source for turbulence. Thus $Ri^{tot}(cr)$ may be substantially larger
than $Ri^{l}(cr)$. This allows the spatial extent of turbulence and
associated mixing for a given gravity wave flux to be larger by about
the same factor \citep{can02}. A combination of these effects could
easily allow the gravity wave mechanism of \citet{gls91} to account
for the mixing in this case. By examining the dissipation from an
energetic standpoint and using a length scale calibrated by fully
non-linear hydro codes with energy sources and sinks, we hope to avoid
this particular difficulty. The \citet{gls91} treatment has certain
advantages. Gravity wave spectra may be dominated by frequencies which
are weakly damped or resonant with characteristic length scales in the
star. We do not initially take this into account. Such a treatment is
necessary for treating angular momentum transport by gravity waves
\citep{ct99,tkz02}, and is likely to be important in the
non-rotational context as well.

\section{TESTS OF THE THEORY}

In this section we present comparisons of models produced by the TYCHO
code incorporating the new convective boundary conditions with
observations in three different evolutionary regimes. This theoretical
description provides useful physical insight into envelope convection
and light element nucleosynthesis, cluster ages and gross stellar
properties including core convection, and complex convection and
advanced nucleosynthesis in evolved stars. No parameter optimization
was used to improve the fit of any models. Two solar models (one with
elment diffusion and one without) were also run as a control, and all
surface observables ($R, T_{eff}, L, X_i$) are in acceptable agreement
with \citet{bpb01}. Errors in the luminosity and $X(^4He)$\ are
consistent with the absence of helium and heavy element settling in
the non-diffusion version of TYCHO. A detailed comparison with
helioseismological constraints on the interior was not performed.
However, the model is in qualitative agreement with suggestions that
the extent of {\it penetrative} convection does not extend much beyond
that predicted by conventional models, while compositional mixing must
go significantly further \citep{bpb01}. The size of the penetrative
convective envelope ($0.727R_{\sun}$, consistent with the no-diffusion
model of \citet{bpb01} and $0.712R_{\sun}$, consistent with the
diffusion model), is similar in TYCHO models with and without the
extra mixing. The slow compositional mixing extends well beyond the
standard convective zone ($\sim 5\times 10^{9}\ {\rm cm}$) when the
new theory is employed.

\subsection{Li and Be in the Hyades and Pleiades}

The burning of lithium and beryllium in pre-main sequence stars
provides a sensitive probe of convective mixing. Lithium is burned at
temperatures above $2.5 \times 10^6$\ K, which can be reached at the
base of convection zones in lower mass stars. A ``lithium edge'' where
the abundance begins to decline from an approximately constant value
is produced at low effective temperatures since the depth of
convection increases with decreasing stellar mass. The location and
steepness of this edge serves to test whether convection in stellar
models reaches as deeply as in real stars. A second dip in the lithium
abundance is seen in F stars ($T_{eff} \sim 6500 - 7000 K$), which
requires additional physics, most likely rotation \citep{thor93, ct99, pt02,
bo02, pin02, tkz02} \citet{gls91} present a gravity wave-excited mixing
treatment which is somewhat consistent with the observational data for the
lithium gap. However, they do not extend the results down to the
lithium edge, so a direct comparison with our work is
difficult. Recent observations have provided similar data for
beryllium, which burns at $3.5 \times 10^6$\ K and thus provides an
additional, related constraint. From a theoretical standpoint, Be
depletions are as simple to estimate as those from Li. Unfortunately,
the atomic transitions of beryllium are located just below the UV
atmospheric cutoff, where ground-based observations of stars with
$T_{eff}$ much below $5500$\ K is difficult. More importantly, at low
$T_{eff}$ a line of magnesium begins to come in strongly almost
on top of the beryllium line, rendering accurate equivalent width
measurements problematic \citep{thor93, pt02, bo02}. The
location of the beryllium edge is therefore not known.

\begin{figure}
\figurenum{2}
\plotone{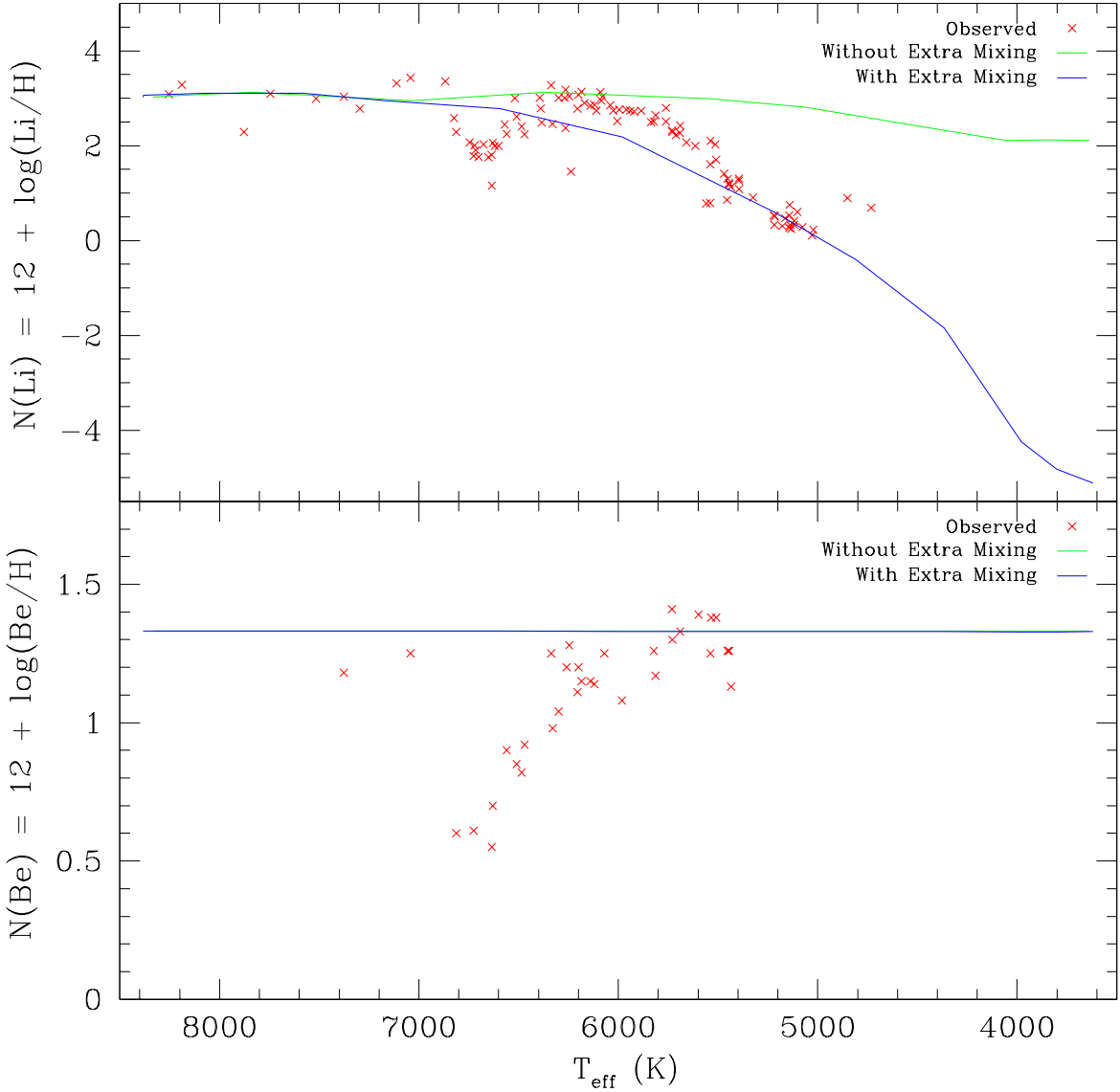}
\caption{Observed Li (top panel) and Be (bottom) abundances from
\citet{thor93, bo02} (crosses, red in the electronic version) along
with calculated values with extra mixing (solid line, blue in the
electronic version) and without (dotted line, green in the electronic
version). The model values with slow mixing follow the observed points
closely at the depletion edge; those without under-predict the
depletion significantly. We suspect that the dips at $T_{eff} \sim
6500 - 7000 K$\ are due to rotation \citep{thor93, ct99, pt02, bo02, tkz02},
indicating that we may be able to separate out the effects of rotation
and dissipative hydrodynamic mixing processes. Note that the dips
appear at approximately the same effective temperature.}
\end{figure}
\placefigure{fig2}

Figure 2 (top) shows calculated surface lithium abundances for stars
of Hyades composition (${\rm [Fe/H] = +0.13\pm0.02}$). Values are
taken at the age of the best fit isochrone for the cluster determined
using photometric data from \citet{db01} and compared with the
observed points from \citet{bo02}. The age of the cluster in our
models is between 650 and 700 Myr, consistent with the age from
conventional overshooting models in \citet{db01}. In our simulations,
the drop-off in lithium with $T_{eff}$ is much too shallow without the
extra slow mixing. Implementing the mixing brings our theoretical
values in line with observations. The lithium dip in F stars is not
reproduced, which is unsurprising as rotation is not included in these
models. We find rather too much depletion of lithium in the models in
the range between the F dip and the depletion edge. We suspect this,
too, is a hallmark of rotational mixing, as in some regimes mixing
appears to actually be {\it damped} by rotation \citep{pt02, kippen,
ch61}. A full calculation of the wave spectrum should also improve the
calculation in this regime. The bottom panel of Figure 2 shows the same
data for beryllium. We find no significant depletion. This is
consistent with observations to the lowest observed $T_{eff}$, and
indicates that our mixing is not excessive. Interestingly, we do not
see any depletion of beryllium at lower effective temperatures. At the
age of the Hyades the lowest mass stars have not finished contracting
onto the main sequence and have not established the deep convective
envelopes necessary to deplete the beryllium. Space-based observations
and data on older clusters could aid in detecting beryllium
depletion. The coincidence in effective temperature between the Li and
Be dips indicates that this is a sensitive test of the depth of the
convective zone. The dip itself may serve as a test of rotation, while
the depletion edge tests non-rotational mixing. 

Simultaneously being able to reproduce Li depletions for clusters of
different ages is problematic for many theories of mixing
\citep{pt02}. In order to test that our description gives a reasonable
time dependence for Li depletion, we modeled the Li edge in the
Pleiades. Figure 3 shows the observed points from \citet{sod93} and
models with the additional mixing. Our models were for our best fit
turn-off age of 120 Myr (see Section 7.2). The models produce somewhat
too much depletion at the lowest $T_{eff}$, but overall the predicted
depletion matches the observations well. The models do not include
molecular hydrogen contributions to the EOS, which becomes significant
at the masses corresponding to the lowest effective temperatures. More
work is required to sort out EOS and opacity effects from the mixing
algorithm in this regime.

\clearpage

\begin{figure}
\figurenum{3}
\plotone{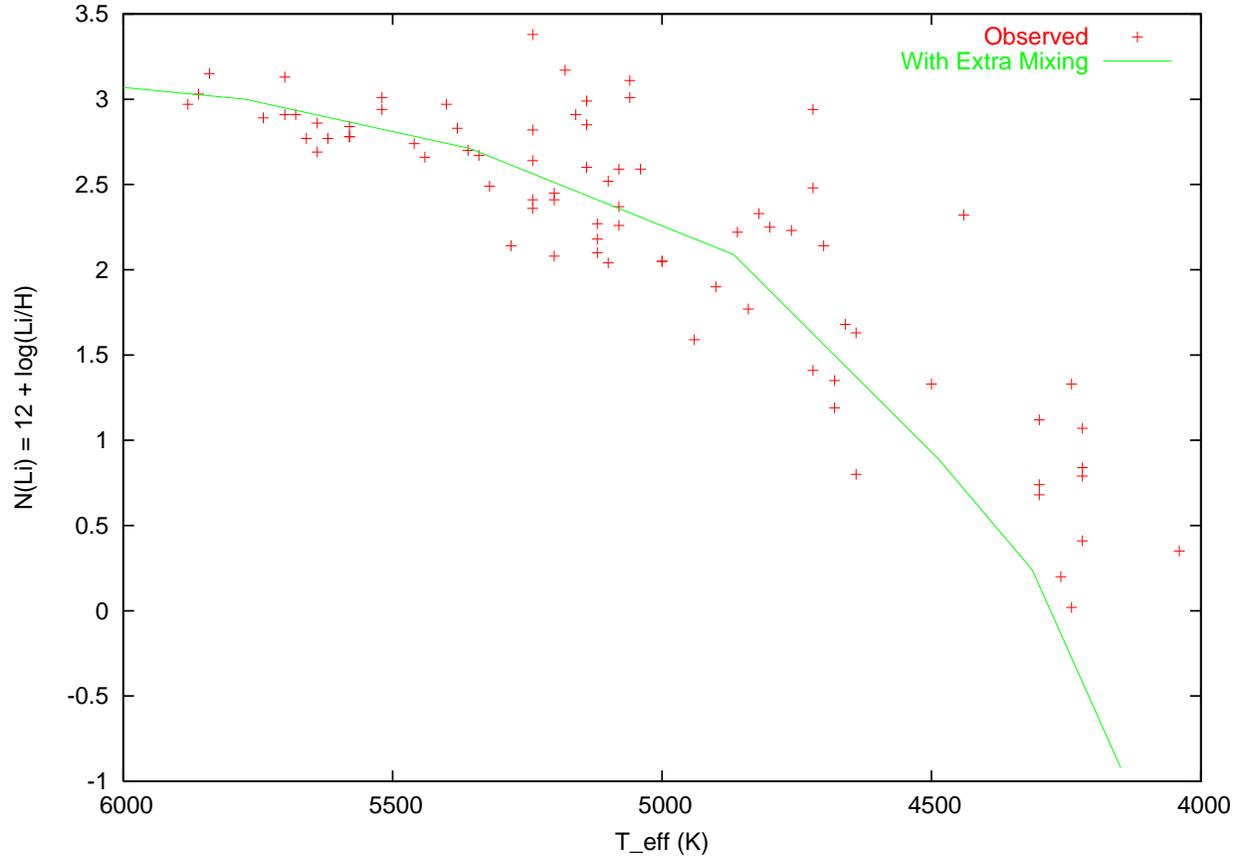}
\caption{Observed Pleiades surface Li abundances (crosses) from
\citet{sod93} plotted with models (solid line) for an age of 120
Myr. The predicted depletions match the observations well except at
the lowest $T_{eff}$. This may be due to an inadequacy in the mixing
model or inaccuracies in the low entropy equation of state.}
\end{figure}
\placefigure{fig3}

\subsection{Comparison With Li Depletion Ages}

The age of the Pleiades has variously been quoted as $75$ to $150$
Myr, with most studies using a value between $75 - 100$ Myr. Recent
determinations of the age using the lithium depletion edge in brown
dwarfs place the age at 125 Myr \citep{ssk98}. Similar uncertainties
exist for other young clusters. Li depletion ages have been determined
for two other clusters, $\alpha$\ Per and IC 2391, with ages of $90\pm
10$\ and $53\pm 5$\ Myr, respectively \citep{aper99, nsp99}. Both ages
are approximately $50\%$\ longer than those derived from conventional
main sequence fitting. Without an independent calibration, it is
equally possible that the Li depletion ages are wrong and turnoff ages
are correct. The depletion ages are, however, consistent with models
with parametrized overshooting calibrated by other methods. In the
absence of further observational constraints we will take the
depletion ages to be a reliable measure. We determine the age of the
clusters by fitting the main sequence turnoff with the extra mixing
included.

\begin{figure}
\figurenum{4}
\epsscale{0.5}
\plotone{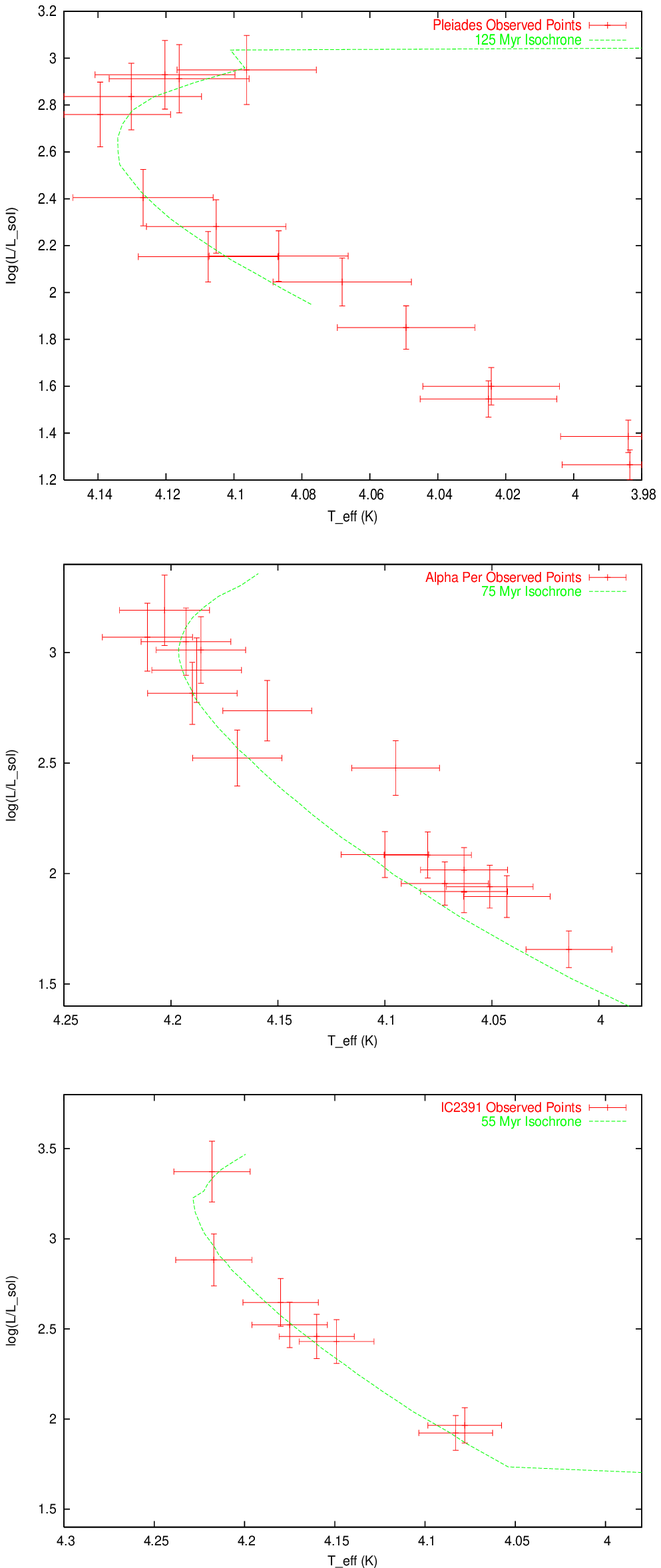}
\epsscale{1.0}
\caption{Observed luminosity and effective temperature for the turnoff
stars of the Pleiades (top), $\alpha$ Persei (center), and IC 2391
(bottom) from \citet{eem}. Crosses are observed stars and lines
represent 120, 75, and 55 Myr isochrones from TYCHO, respectively. The
error bars are representative, and do not properly take into account
systematic errors. The isochrones are a reasonable fit at the lithium
depletion ages of the clusters without recourse to parameter
optimization.}
\end{figure}
\placefigure{fig4}

Models were run for masses from $3.0$ to $ 6.0$ \sol\ in increments of
$0.1$ \sol . The models were run at solar metallicity, which is within
the error bars for the observations \citep{bf90, ran01}. The $L$\ and
$T_{eff}$\ conversions from observational data are taken from
\citet{eem}. Figure 4 shows our isochrones for the Pleiades (top),
$\alpha$ Persei (center), and IC 2391 (bottom) at 120, 75, and 55 Myr,
along with observed values corrected for differential reddening across
the clusters. The error bars on the observations are sufficiently
large that further refinement of the age was not attempted. The
turnoff ages with the extra mixing are 120 Myr for the Pleiades, 75
Myr for $\alpha$\ Per, and 55 Myr for IC 2391, consistent with the
ages determined from lithium depletion in brown dwarfs.

One additional constraint is also reproduced. There is one white dwarf
member of the Pleiades with a mass of $\sim 1$\  \sol\  \citep{weg91}. Our
models produce a white dwarf progenitor of $\sim 1$\  \sol\  at the age of
the cluster from an initial mass of $\sim 5.5$\ \sol .

\subsection{Carbon Stars}

The term ``carbon star'' is variously used to describe a menagerie of
objects with surface abundances of carbon enhanced relative to
oxygen. The group includes evolved stars on the AGB, subsets of white
dwarfs and Wolf-Rayets, and cool dwarfs. Only the first
category will be discussed here. The evolved stars further may show
enhancements in s-process elements and lithium. Observations of
$^{99}{\rm Tc}$, which has a half life of $2 \times \ 10^5$\ years,
indicates that the products of {\it in situ} nuclear processing are
being mixed to the surface. The s-process elements and enhancements of
lithium and $^{13}{\rm C}$\ require burning in a region enriched in
both protons and the products of triple $\alpha$\ burning
\citep{wk98,cf71}. This is difficult to reproduce with traditional
stellar evolution codes, since the products of partial triple $\alpha$
burning are not in general mixed into hydrogen burning regions or
further to the surface \citep{bus99}.

Making comparisons between models and observed carbon stars is
difficult, as the class includes such a large variety of stars. The
masses of carbon stars for low metallicity populations appear to range
from $\sim 0.8$\ to $\sim 6$\  or $8$\ \sol . Absolute bolometric
magnitudes ranging from ${\rm M_{bol}} = 0$ to $-8$ ($L/L_{\sun} \sim
10^2 - 10^5$), $T_{eff} \sim 2000 - 5000 {\rm K}$, and radii from
approximately 2.4 - 2.7 AU \citep{w73,wk98}. Carbon stars appear to
come in a range of metallicities, but the ratio of C to M stars
increases greatly from the Galactic bulge to the Magellanic
Clouds. There is a definite trend toward increasing efficiency of
carbon star production at low metallicities \citep{bl80}.

Implementing the present theoretical description of convective
boundary conditions in TYCHO, we obtain carbon stars without further
modification of the code. In light of the variety inherent in the
class, this does not, by itself, demonstrate much about the
effectiveness of the treatment, but when considered along with the
success in a range of other regimes, is a promising sign. Exact
isotope ratios are dependent not only on the boundary conditions, but
also on the time dependent treatment of the compositional mixing
inside the convective region itself. A subsequent paper will examine
CNO and s-process nucleosynthesis for a range of masses and
compositions.

We find that a 6 \sol\ star at $z = 0.001$ produces a star with
surface carbon in excess of oxygen at the beginning of the thermal
pulse AGB. The luminosity and $T_{eff}$ are consistent with observed
quantities for C-N stars. Carbon approaches but never exceeds oxygen
for a solar metallicity model, as we might expect from the observed
bias toward low metallicity environments. Figure 5 shows the surface
abundances of $^{7}{\rm Li}$, $^{12}{\rm C}$, $^{13}{\rm C}$,
$^{14}{\rm N}$, and $^{16}{\rm O}$\ for the final $10^5$ years of the
model track. The star shows a pulse of elevated lithium and $^{13}{\rm
C}/^{12}{\rm C}$ ratio at the beginning of the carbon star phase. This
is consistent with the (again, wide range of) observed values for C-N
stars, which show a bimodal distribution in carbon isotope ratios and
enhanced lithium values.  This may reflect an evolutionary trend
\citep{wk98}.

\begin{figure}
\figurenum{5}
\plotone{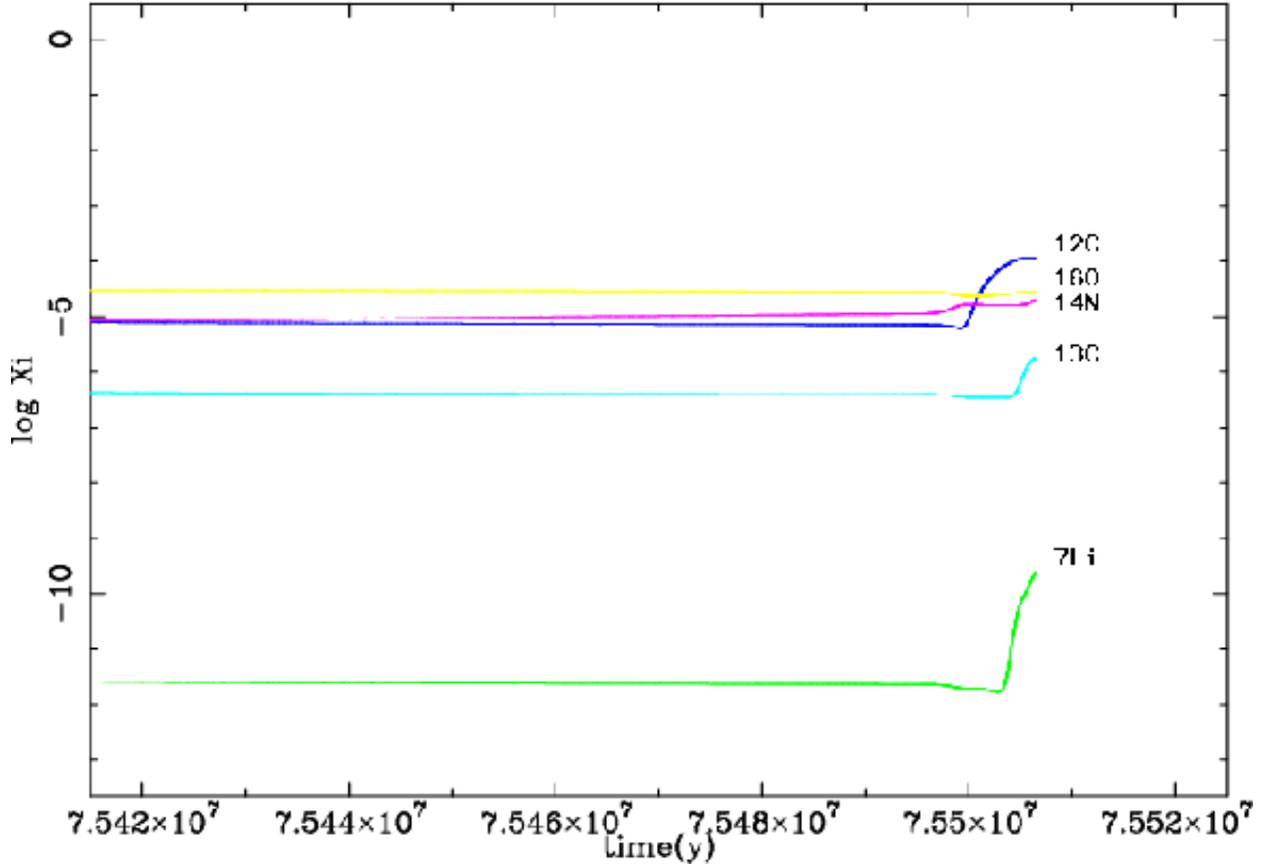}
\caption{Surface abundances of $^{7}{\rm Li}$, $^{12}{\rm C}$,
$^{13}{\rm C}$, $^{14}{\rm N}$, and $^{16}{\rm O}$\ for the final
$10^5$ years of a $z = 0.001$ 6 \sol\ model evolutionary track,
corresponding to the beginning of the thermal pulse AGB. The carbon
abundance has exceeded oxygen at the surface and is accompanied by a
pulse of $^{7}{\rm Li}$\ and $^{13}{\rm C}$. }
\end{figure}
\placefigure{fig5}

\section{CONCLUSIONS}
We take a novel approach to the problem of mixing in stars by
identifying phenomena which emerge in nonlocal, nonlinear,
multi-dimensional hydro simulations. These simulations appear to
successfully reproduce behavior on the large scale which transport
most of the flux of energy and material. We then develop a theoretical
description of this large scale behavior. This facilitates the
transition from observed phenomenology to a predictive understanding
which can be of use in the wider context of stellar evolution.

Several fundamental, if not surprising, results arise from
implementing such a physical theory. First, the boundary between
convectively stable and unstable regions cannot be treated as
spherical or static, even in a one-dimensional approximation of the
sort necessary for stellar evolution calculations. Hydrodynamic
processes seen in multiple dimensions must be taken into
account. Second, a careful treatment of the boundary conditions always
results in extra mixing beyond the formal boundary. Third, a single
physical process operates in both core and surface convective
zones. Fourth, implementation of this theory in the stellar evolution
code TYCHO contributes significantly to solving problems in several
different regimes of stellar evolution. This is accomplished with only
one parameter that does not fall directly out of the theoretical
description, namely the dominant wavelength of the gravity waves
driving the slow circulation in the radiative zone. Even this
parameter is (a) a quantity with physical meaning, and (b) not allowed
to vary, being fixed by data from numerical simulations. Finally, if
this model continues to be as successful as it has thus far at
explaining non-rotationally induced mixing, it will allow us to
isolate the rotational contribution to stellar physics with a fair
degree of confidence.

We reproduce the Li depletion edge in the Hyades and Pleiades. We find
cluster ages for three young clusters consistent with ages determined
from measurements of Li in brown dwarfs and for the Hyades as
determined by main-sequence fitting with alpha-overshoot. The theory
also generates reasonable carbon star models on the AGB. We expect
that the physics and nucleosynthetic yields of supernovae and gamma
ray bursts may be sensitive to the rotational properties of the star,
core sizes, and final composition profiles at core collapse. It is
essential to produce accurate initial models in order to generate
realistic models of the explosion. This requires a physical, rather
than simply phenomenological, characterization of the hydrodynamic
mixing and rotation in stars. These factors also influence chemical
enrichment from AGB stars and thermonuclear supernovae. These results
may significantly improve our understanding of these processes, which
impact issues as disparate as cluster ages, and thus timescales
observed for disk evolution in pre-Main Sequence stars, to the
nucleosynthetic history of the universe.

We stress that this result is merely a first step toward completely
and predictively characterizing the mixing in stars. Numerical
simulations have already illuminated physical processes which have
changed our understanding of stellar astrophysics. Experiments with
higher resolution, more complete physics, and a wider variety of
geometries and thermodynamic conditions appropriate to the range
encountered in stars are vital, as they may well display yet more
complex phenomena.  Several other processes remain to be integrated
into a complete picture of stellar mixing. This treatment does not
take into account the effect of magnetic fields, which provide an
upwardly biased buoyancy force and, when overlapping the convective
boundary, coupling between stable and unstable fluids. Coupling
between rotation and convective fluid motions must also be
considered. Finally, changes to the nuclear burning and convection
resulting from the ingestion of fresh fuel into a convective core or
shell must be more carefully explored. We are confident, however, that
the careful treatment of stellar hydrodynamics in both convective and
radiative regions, plays an essential role in understanding the
important problem of mixing in stars.

\clearpage

\end{document}